# Survey on Sparse Coded Features for Content Based Face Image Retrieval

D. JohnVictor[1], G. Selvavinayagam[2]

[1](PG Student, Department of Information Technology, SNS College of Technology, Coimbatore, Tamil Nadu, India)

[2](Assistant Professor, Department of Information Technology, SNS College of Technology, Coimbatore, Tamil Nadu, India)

**ABSTRACT:** Content based image retrieval, a technique which uses visual contents of image to search images from large scale image databases according to users' interests. This paper provides a comprehensive survey on recent technology used in the area of content based face image retrieval. Nowadays digital devices and photo sharing sites are getting more popularity, large human face photos are available in database. Multiple types of facial features are used to represent discriminality on large scale human facial image database. Searching and mining of facial images are challenging problems and important research issues. Sparse representation on features provides significant improvement in indexing related images to query image.

## 1. INTRODUCTION

Image retrieval system usually use low level features (e.g., color, texture) to represent face images. We consider the problem of retrieving human faces from using low level features with varying expression and illumination, as well as occlusion and disguise. New theory from sparse signal representation offers the key to addressing this problem. If sparsity on facial features is properly addressed, the choice of features is no longer critical. Various image feature descriptors such as Local Binary Pattern (LBP), Scale Invariant Feature Transform (SIFT) are used to represent images. We first convert each feature descriptor into a sparse code, and aggregate each type of sparse coded features into a single vector. Multiple vectors from different types of features are then concatenated to obtain the final representation. This approach allows adding more types of features to improve discriminability without scalability issues. Invariant features using the sparse-coding framework, outperforms the state of the art both in retrieval performance and in scalability. Basic image retrieval system works on calculating the Euclidean distance between all pairs of faces of two images. The smallest of these distances is then interpreted as similar faces and these images are retrieved.

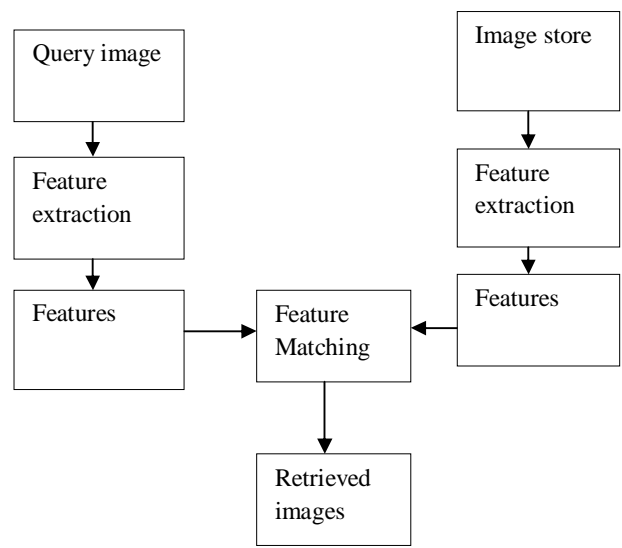

figure 1 content based image retrieval system

This type of system affects by its low level semantics. Sparse representation provides solutions to this semantic loss by using improved feature descriptors.

## 2. FACE RECOGNITION VIA SPARSE REPRESENTATION

Automatic human face recognition has problem with varying expression and illumination. In this paper [1] general classification algorithm has been proposed with sparse representation to recognize faces. Extensive studies in human and computer vision about the effectiveness of partial features in recovering the identity of a human face are discussed in this paper. Extensive studies in human L1 minimization problem used to solve sparse representation used for classifying and validating face image samples. This new classification framework addresses two important issues: 1.feature extraction 2.robustness to occlusion. Sparse representation role in feature extraction is which low level features are more informative to recognize object that might be extracted. Occlusion poses are obstacles to face recognition in real world images. The subsequent sparse representation on occluded test image with respect to dictionary selection of trained images obviously separates occlusion components from test image. Proposed framework can tolerate in face recognition on occluded images with source and error separation algorithm. Partition the image into blocks and process each block independently can improve face recognition performance with robust to occlusion. This robustness allows algorithm to be more tolerant to pose and illumination.

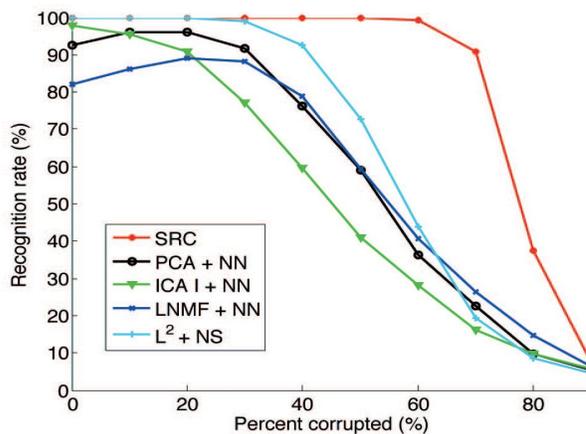

figure 2 the recognition rate across the entire range of corruption for various algorithms

In future, framework needs to integrate with feature matching techniques and nonlinear deformation models to reduce distribution of face image under varying pose

## 3. SPARSENESS OF VISUAL WORDS

Challenging problem in image object retrieval is targeting object only covers small portion image. Traditional Bag-of-Words (BoW) quantizes low level features into Visual Words (VW) and fail to address problems relate to variations in lighting and viewpoints. In this paper [2] proposed system mine auxiliary visual words by combining both visual and textual information. This system investigating variant optimization methods to discovers auxiliary visual words. Graph of images constructed by visual and textual information. Auxiliary visual words are extracted from this visual and textual graph. New retrieval system includes Scale Invariant Feature Transform (SIFT) and Bag-of–words model. SIFT descriptors quantize features into visual words. BoW adopts k-means clustering to construct vocabulary. This approach shows that the occasion of visual words in different images is very sparse. Experimental results on two image databases of different sizes shows similar images might have few common VWs. Visual words are low level features and lack of semantics meaning.

Retrieval system often fails to retrieve image with different view angle. This problem can be addressed by using textual information on images to improve semantics in sparse coding. MapReduce technique used to cluster images on the image graph by applying Affinity Propagation (AP). Images in the same visual cluster are visually similar to each other and represented by VWs in each visual cluster. Propagation of visual words is important to obtain different VWs. These too many VWs and decrease the precision. Removing those irrelevant or noisy VWs, proposed approach select those representative VWs in each visual cluster. Thus the proposed system greatly improves retrieval performance compared to BoW model. In future to preserve sparseness for visual words, L2 loss L1 normalization problem solvers can be used to maximize its accuracy and efficiency.

## 4. SPARSE CODING FOR IMAGE CLASSIFICATION AND RETRIEVAL

Sparse codes of local features in an image at multiple spatial scales represents image with more accuracy by using locality constraints.. In this paper [3] new algorithm proposed to design dictionaries for sparse coding of image descriptors and learned dictionaries are used to recognize objects. Kernel Local Sparse Coding (K-LSC) methods used to find similarities of features and Radial Basis Function (RBF) kernel used to learn dictionaries. In this system supervised local feature sparse coding methods developed using sub image heterogeneous feature for retrieving face image. Finite collection of normalized features is referred to as a dictionary. Sparse coding based dictionary learned approach provide superior performance in many applications. Proposed system use optimization problem to solve dictionary learning and sparse coding. K-SVD algorithm used to solve the problem of dictionary learning is a generalization of data clustering. Patches are extracted from raw images can be directly employed for classification. Bag-of- words or the state-of-the-art Spatial Pyramid Matching (SPM) technique aggregate sparse codes obtained from the facial features. Sparsity is regularized to promote discrimination. Discriminative frameworks such as the linear discriminant analysis and linear Support Vector Machines (SVM) can be incorporated in to obtain sparse codes for classification tasks. Features extracted from the images are either Scale Invariant Feature Transform (SIFT) descriptors or Local Binary Pattern (LBP) descriptors. These feature descriptors are coded using vector quantization with a K-means dictionary.

TABLE 1
Comparisons of Classification Accuracy

| Method | Accuracy % |
|---|---|
| Sparse Coding | 80.28 |
| Kernel Sparse Coding | 83.08 |
| Laplacian SC | 89.75 |
| Locality-Constraint Linear Coding | 89.78 |
| Local SC | 89.94 |

Supervised information incorporating into local sparse coding will result in highly discriminative features for retrieval. Simulations demonstrated the gain in image retrieval performance obtained by performing supervised coding. Incorporating supervised information into local sparse coding will provide improved precision-recall rates.

## 5. SPARSE CODING WITH IDENTITY CONSTRAINTS

Face retrieval task is more related to face recognition. Face recognition system uses texture descriptors to prove better recognition performance. Although descriptors used in face recognition can be applied to face retrieval. Creating index structures on descriptions with high dimensionality is not a trivial task. In traditional image retrieval system such as Bag-of-Words (BoW) model created by local feature descriptors does not provide better performance. It is because face images have higher intra-class variance and these methods neglect the important spatial information in face image. This paper [4] proposed a face image retrieval system using identity based information to solve this problem. Most of database images are including identity information, gender information, social network, etc so it is easy to have its necessity. Proper usage of such information in the retrieval system improves the results. Sparse coding has shown promising results on many different applications such as image denoising and classification. A novel coding scheme refines the result of sparse coding based on identity information. Sparse coding with identity constraint achieves 70% relative improvement in Mean Average Precision over the baselines. In future we can boost up performance by combining high level semantics features into sparse code generation framework.

## 6. ATTRIBUTE ENHANCED SPARSE CODEWORDS

Existing face image retrieval system use low level facial features to represent face image. But these low level features are lack of semantics meaning and affects retrieval performance because of facial images have high inter class variations. Face images of different people might be very close in the low-level feature space. In this paper [5], proposed system utilized high level human facial attributes into face image representation and index structure Human facial attributes (e.g., hair, age, gender, personal, race) are provide high-level semantic descriptions about a human face. By combining low-level features with high-level human attributes can provides better feature representations. Because certain people might have similar attributes it loses discriminability among too many face images in database.

To address these problems two orthogonal methods are proposed: 1. Attribute-enhanced sparse coding 2. Attribute-embedded inverted indexing. Attribute-enhanced sparse coding uses facial attributes combined with texture features to construct semantic codewords.

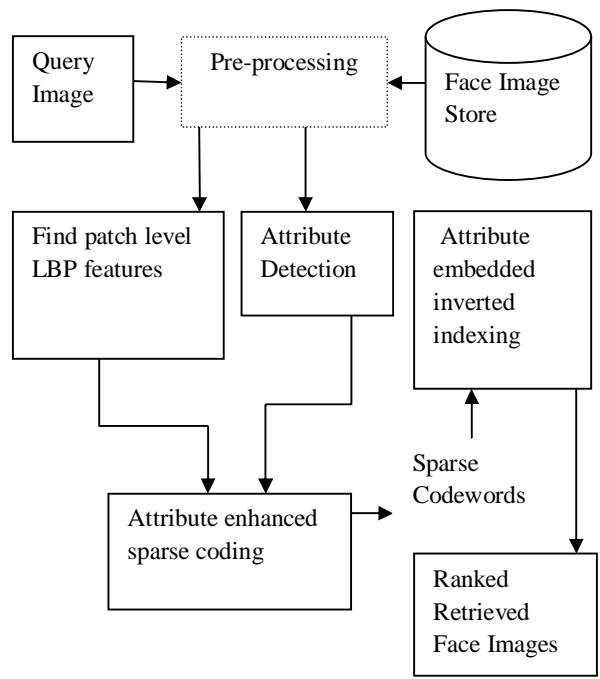

figure 3 face image retrieval with sparse coding

Attribute-embedded inverted indexing creates inverted index structure by incorporating binary attribute signature in addition to the sparse codewords. Certain facial attributes are not correlated to human identity. So these attributes are degrading retrieval performance.

## 7. SPATIAL PYRAMID MATCHING USING SPARSE CODING

Spatial Pyramid Matching (SPM) kernel is highly successful in image classification. In this paper [6] extensive SPM Method has been developed by generalizing Vector Quantization (VQ) to Sparse Coding (SC). Both Bag-of–Features (BoF) and SPM must be applied together with a particular type of nonlinear Mercer kernels, e.g. the intersection kernel or the Chi-square kernel. The bag-of-words approach to image classification computes such a histogram for each image represented by an unordered set of local descriptors. But in extensive SPM approach, the image's spatial pyramid histogram representation is a concatenation of local histograms in various partitions of different scales. SC used to derive image representations because of its following advantages. 1. SC coding can achieve lower reconstruction error due to the less restrictive constraint compared to VQ. 2. Sparsity captures salient properties of images. 3. Image statistics research clearly reveals that image patches are sparse signals.

Sparse coding with SPM (ScSPM) based on linear kernel achieves better performance on different datasets and does not require much more computation to mention that the nonlinear methods require much more computation. This system taking advantages of compatibility of linear models with SIFT sparse codes. Frame-based approach in human face detector and detects events of interest employed to reduce computation effort. Related works shows using Feed-forward network can accelerate sparse coding by reducing time taken on encode SIFT feature descriptions.

## 8. CONCLUSION

Using sparse coding we can obtain discriminative image representation that outperforms state of art. In this survey only few sparse coding techniques achievements in content base face image retrieval systems are reviewed. This paper provides comprehensive survey on facial feature representation, clustering of visual words, reducing semantics gap and retrieval performance. With these various approaches Attribute-enhanced sparse coding outperforms when compared to other approaches because of its high level descriptions on facial features. By introducing new techniques in future will improve semantics representation on sparse codewords that provide significant retrieval performance.

## REFERENCES

[1] J. Wright, A. Yang, A. Ganesh, S. Sastry, and Y. Ma, "Robust face recognition via sparse representation," *IEEE Trans. Pattern Anal. Mach. Intell., vol. 31, no. 2, pp. 210–227*, Feb. 2009.

[2] B.-C. Chen, Y.-H. Kuo, Y.-Y. Chen, K.-Y. Chu, and W. Hsu, "Semi-supervised face image retrieval using sparse coding with identity constraint*," in Proc. ACM Multimedia*, 2011.

[3] Jayaraman J. Thiagarajan, Karthikeyan Natesan Ramamurthy, Andreas Spanias," Local Sparse Coding for Image Classication and Retrieval", *Pattern Recognition Letters*, may 2012.

[4] Y.-H. Kuo, H.-T. Lin, W.-H. Cheng, Y.-H. Yang, and W. H. Hsu, "Unsupervised auxiliary visual words discovery for large-scale image object retrieval," *in Proc. IEEE Conf. Computer Vision and Pattern Recognit.,* 2011.

[5] Bor-Chun Chen, Yan-Ying Chen, Yin-Hsi Kuo, and Winston H. Hsu, "Scalable Face Image Retrieval Using Attribute-Enhanced Sparse Codewords", *IEEE Transactions On Multimedia, VOL. 15, NO. 5* AUGUST 2013.

[6] J. Yang, K. Yu, Y. Gong, and T. Huang, "Linear spatial pyramid matching using sparse coding for image classification", *CVPR*, 2009.